# Secure Fractal Image Coding

Shiguo Lian

France Telecom R&D Beijing

2 Science Institute South Rd., Haidian District

Beijing, 100080, P.R China

E-mail: shiguo.lian@orange-ftgroup.com; sg_lian@163.com

**Abstract.** In recent work, various fractal image coding methods are reported, which adopt the self-similarity of images to compress the size of images. However, till now, no solutions for the security of fractal encoded images have been provided. In this paper, a secure fractal image coding scheme is proposed and evaluated, which encrypts some of the fractal parameters during fractal encoding, and thus, produces the encrypted and encoded image. The encrypted image can only be recovered by the correct key. To keep secure and efficient, only the suitable parameters are selected and encrypted through investigating the properties of various fractal parameters, including parameter space, parameter distribution and parameter sensitivity. The encryption process does not change the file format, keeps secure in perception, and costs little time or computational resources. These properties make it suitable for secure image encoding or transmission.

# 1 Introduction

With the development of multimedia technology and Internet technology, multimedia data are used more and more widely in human's life. Taking images for example, they can bring more information than text or binary data, e.g., advertisements, pictures, maps, etc. In some applications, the sensitive images related to politics or commerce, such as military maps, medical images or multi-resolution images, should be protected in order to avoid unauthorized users knowing it. This demand activates the research on image encryption. Image encryption technique [1][2] transforms the image into an unintelligible form under the control of a key. Only the user having the correct key can recover the image. Till now, various image encryption algorithms have been reported, which can be classified into two types, i.e., raw image encryption and compressed image encryption.

In raw image encryption, the image is encrypted before compression. The simplest method is image permutation [3,4] that changes the pixels' position randomly. Recently, chaos-based ciphers are often used in image encryption. For example, the cascaded chaos maps are used to construct the stream cipher for image encryption [5], the discrete Kolmogorov flow map is used to design the parallel image encryption algorithm [6], two chaotic maps are combined to shuffle the image pixels [7], nonlinear chaotic algorithm is introduced to replace linear function in image encryption [8], discrete exponential chaotic maps' confusion and diffusion properties are improved and used to design the image encryption algorithm [9], the Chaotic Neural Network is used to design the stream cipher for images [10], the 2-dimensional Baker map is used to construct the block cipher for images [11,12,13,14], the 3-dimensional chaotic maps are used to make the block ciphers for images [15,16]. These ciphers encrypt the uncompressed images directly without considering of compression.

In practice, images or videos are often compressed in order to save the cost of storage space or transmission loading. Thus, it is more reasonable to encrypt the compressed data. Furthermore, considering that images or videos are often of large volumes, encrypting the compressed data completely will cost much time. Thus, it will reduce the computational cost if encrypting only part of the compressed data. For example, the DCT coefficients' signs are encrypted in DCT transformed blocks [17], the data blocks are permuted in frequency domain [18,19], and both the coefficients' signs and block positions are encrypted [20,21]. These algorithms encrypt only some parameters in the image, reduce the encrypted data volumes, and thus, improve the encryption efficiency. The key problem is how to select the parameters. Considering that different compression method produces different parameters, different encryption algorithm should be designed for different compression method.

Recently, fractal image coding [22,23] attracts more and more researchers, which adopts the self-similarity in images to compress image data. Generally, in these compression methods, the image is partitioned into blocks. According to the property of self-similarity, for each block, a fractal transformation can be obtained, which is determined by some fractal parameters. Thus, for each block, only the corresponding fractal parameters are stored instead of the block data themselves, which reduces the data size. Most of the recent works [24,25,26] focus on the method to obtain the suitable fractal transformation.

Till now, there is no solution for image encryption in fractal coding. This paper aims to give an encryption scheme for fractal image coding. Firstly, the fractal parameters' properties are investigated, such as the parameter space, parameter distribution and parameter sensitivity. Then, based on the investigation, the suitable fractal parameters are selected, and the secure fractal image coding scheme is constructed, which combines encryption/decryption operation with fractal image encoding/decoding process. Finally, some experiments and analyses are given to show the proposed scheme's performances, such as security and efficiency.

The rest of the paper is arranged as follows. In Section 2, fractal image coding is briefly introduced. Then, the secure fractal image coding scheme is presented in Section 3. And the performances, including security and efficiency, are evaluated in Section 4. Finally, some conclusions are drawn and future work is presented in Section 5.

## 2 About Fractal Image Coding

Fractal image coding [27] adopts the self-similarity in an image to find the iterated contract transform for each block and stores only the parameters of contract transform instead of the image pixels. The first fractal image coding method is proposed by Jacquin [28,29], which partitions the image into squared domain blocks and range blocks, finds the most matched domain block corresponding to each range block, determines the suitable contract transform that makes the transformed domain block most similar to the original range block, and stores the parameters of the contract transform. Most of existing works focus on the method to obtain the most suitable contract transforms. For example, Thomas and Deravi [30] uses region-growing method to combine some range blocks and makes the range blocks more adaptive with image content, Franich [31] partitions the image into range blocks with a quadtree-based method in order to make the encoding process adaptive with the image content, Truong et al. [23] utilizes the spatial correlations in both the domain blocks and the range blocks to reduce the searching space during block matching, He et al. [24] uses the one-norm of normalized block to avoid the excessive search in block matching, and Zhou et al. [22] uses a special unified feature (UFC) to reduce the search space and borrows DCT coder to improve the quality of the reconstructed image.

### 2.1 Iterated Contract Transform

In fractal image coding, for each range block, the iterated contract transform, also named fractal transform, is determined. Set $M$ the matrix space of the image, $\delta$ the given metric of matrix error, $\mu_{orig}$ the original image to be encoded and $\tau$ the contract transform. The contract transform $\tau$ should make the image contracted from $M$ to $\mu_{orig}$ after iterated transforms. Generally, $\tau$ satisfies the following conditions [29].

$$\exists s < 1, \forall \mu, \nu \in M, \delta(\tau(\mu), \tau(\nu)) \leq s\delta(\mu, \nu), \tag{1}$$

and $\delta(\mu_{orig}, \tau(\mu_{orig}))$ is small. Here, $s$ is named the contract factor of the contract transform $\tau$. Since $\tau$ often needs much smaller storage space than $\mu_{orig}$ does, $\tau$ can be regarded as the compressed form of the original image $\mu_{orig}$.

**2.2 Block Based Implementation**

Block based fractal image coding [29,30,31,32] is now the most popular method. Generally, the original image $\mu_{orig}$ is partitioned into domain blocks and range blocks. For each range block, the most similar domain block will be found and the fractal transform is determined. Taking N range blocks for example, the encoding process is to find a set of fractal transforms $\{\tau_1, \tau_2, \cdots, \tau_N\}$ that can transform an arbitrary image $\mu_0$ into the original image $\mu_{orig}$ after iterations. In practical implementation, fractal transform $\tau_i$ (i=0,1,…,N-1) is composed of geometric transform $S_i$ and other transform $T_i$. Among them, geometric transform includes the size scale and block position, while the other transform includes contrast scaling $\alpha_i$, luminance offset $\Delta g_i$ and isometric transformation $\ell_i$. Here, $\alpha_i$ ranges in [0,1] in order to keep the fractal transform contracted, $\Delta g_i$ is the difference between pixels in the range block and the ones in the transformed domain block, and $\ell_i$ includes the rotation and symmetric transform in horizontal, vertical or diagonal direction. Thus, for the i-th range block, the fractal transform $\tau_i$ is defined as

$$\tau_i = T_i \circ S_i. \tag{2}$$

Set $R_i$ a range block and $D_i$ the corresponding domain block, then

$$\widetilde{R}_i = \tau_i(D_i) = T_i(S_i(D_i)) = \ell_i(\alpha_i(S_i(D_i)) + \Delta g_i). \tag{3}$$

Here, $\widetilde{R}_i$ is the approximate value of $R_i$ estimated by $D_i$.

The decoding process is composed of iterated contract transforms. That is, for arbitrary initial image $\mu_0$, each of its range block is transformed by the iterated fractal transform, which makes the initial image $\mu_0$ contracted to the original image $\mu_{orig}$.

**2.3 The General Implementation**

In fractal image coding, the parameters of fractal transform are stored, which include the position of the matched domain block $D_{xy}$ ($D_x$ in horizontal direction and $D_y$ in vertical direction), contrast scaling $\alpha_i$, luminance offset $\Delta g_i$ and isometric transformation $\ell_i$, etc [32]. Additionally, if the block partitioning is based on quadtree method, then the height information of the quadtree should also be stored. The position information is in relation with the number of domain blocks. For example, if the number of partitioned domain block is 128x128, then the position information is of 14 bits (7 bits for $D_x$, and 7 bits for $D_y$). For the parameter $\alpha_i$, the more the assigned bits, the higher the encoded image's quality. For the parameter $\Delta g_i$, the more the bits, the higher the encoded image's quality. Generally, the number of bits for $\Delta g_i$ is in relation with the number of pixel levels. For example, if the image is of 256 colors, then $\Delta g_i$ is of 9 bits (one bit for the sign). The parameter $\ell_i$ is often of 3 bits, since there are often 8 kinds of isometric transforms, as shown in Table 1. Thus, the typical assignment for fractal parameters is listed in Table 2.

**Table 1.** Cases of isometric transform $\ell_i$

| No | Transform | Bits |
|----|-----------|------|
| 1 | unchanged | 000 |
| 2 | Symmetric transform in the middle horizontal axis | 001 |
| 3 | Symmetric transform in the middle vertical axis | 010 |
| 4 | Symmetric transform in the first diagonal line | 011 |
| 5 | Symmetric transform in the second diagonal line | 100 |
| 6 | Rotate $90^0$ | 101 |
| 7 | Rotate $180^0$ | 110 |
| 8 | Rotate $-90^0$ | 111 |

Table 2. Bit number of fractal parameters

| Parameter | Number of bits |
|---|---|
| $D_{xy}$ | 7+7 |
| $\alpha_i$ | 6 |
| $\Delta g_i$ | 8+1 |
| $\ell_i$ | 3 |
| Sum | 32 |

**3 Secure Fractal Image Coding Scheme**

In the proposed secure fractal coding scheme, some sensitive parameters in fractal image coding will be selected and encrypted. In the following content, the principles of secure partial encryption will be introduced, the properties of the parameters of fractal image coding will be investigated, and the suitable parameters are selected and encrypted with secure ciphers.

**3.1 Principles of Secure Partial Encryption**

In partial encryption, only some parameters of image data are encrypted while the others are left unchanged. To keep secure, some principles [2] are required to select the suitable parameters:

i) The parameter with the properties of large space and random distribution is preferred to be encrypted. The larger the parameter space is, the more difficult the brute-force attack [33] is. If the parameter is in random distribution, then the difficulty of statistical attack will be increased.

ii) The parameter with high sensitivity is preferred to be encrypted. The parameter sensitivity denotes the quality degradation caused by a unit error in the parameter. If the quality degradation is big, the parameter is regarded as of high sensitivity. Encrypting the parameter with high sensitivity is easy to make the image content unintelligible.

iii) For parameter encryption, the cipher with high security is preferred. The parameter should be encrypted with the cipher that has high key sensitivity. It can confirm that a slight difference in the key will lead to great differences in the parameter, and thus, a slight different key will make the decryption failed.

### 3.2 Investigation of Fractal Parameters

#### 3.2.1 Parameter Space

In fractal image coding, different parameter has different space. Under the condition of knowing only the encrypted parameter, the parameter space determines the brute-force times for guessing the original parameter. According to the example shown in Table 2, the parameter spaces are listed in Table 3.

**Table 3.** Parameter spaces

| Parameter - a | Parameter space – S(a) |
|---|---|
| $D_{xy}$ | $2^{14}$ |
| $\alpha_i$ | $2^6$ |
| $\Delta g_i$ | $2^9$ |
| $\ell_i$ | $2^3$ |

As can be seen, $\Delta g_i$ and $D_{xy}$ have larger parameter spaces than $\alpha_i$ and $\ell_i$. Additionally, the parameter space of $D_{xy}$ is in relation with the number of domain blocks. The more the domain blocks, the larger the space of $D_{xy}$. Otherwise, on the contrary. Among these parameters, the rotation parameter $\ell_i$ has the limited space that may be fragile to brute-force attacks, and thus, it is not suitable for encryption.

#### 3.2.2 Parameter Distribution

In fractal image coding, for different range block, a parameter may get different value from its parameter space. By computing the frequency of each parameter value, we can get the parameter's distribution. Suppose the image is partitioned into M×M domain blocks, the happen probability of each parameter value is defined as

$$P(a,j) = \frac{F(a,j)}{S(a)}. \tag{4}$$

Here, a (a is $\Delta g_i$, $D_{xy}$, $\alpha_i$ or $\ell_i$) is the fractal parameter, S(a) is the parameter space of a, P(a,j) is the probability of the parameter a that satisfies a=j (j=0,1,…,S(a)-1), F(a,j) is the frequency of parameter a when a=j (j=0,1,…,S(a)-1).

Suppose the image is partitioned into 128×128 range blocks and 127×127 domain blocks and the parameter spaces of $\Delta g_i$, $D_{xy}$, $\alpha_i$ or $\ell_i$ is same to the one proposed in Table 3. Taking five images (256x256 Lena, 512x512 Boats, 512x512 Eltoro, 256x256 Village and 256x256 Scene) for example, we test and compute the probability of each fractal parameter. For each parameter, the average probability between the 5 images is shown in Fig. 1.

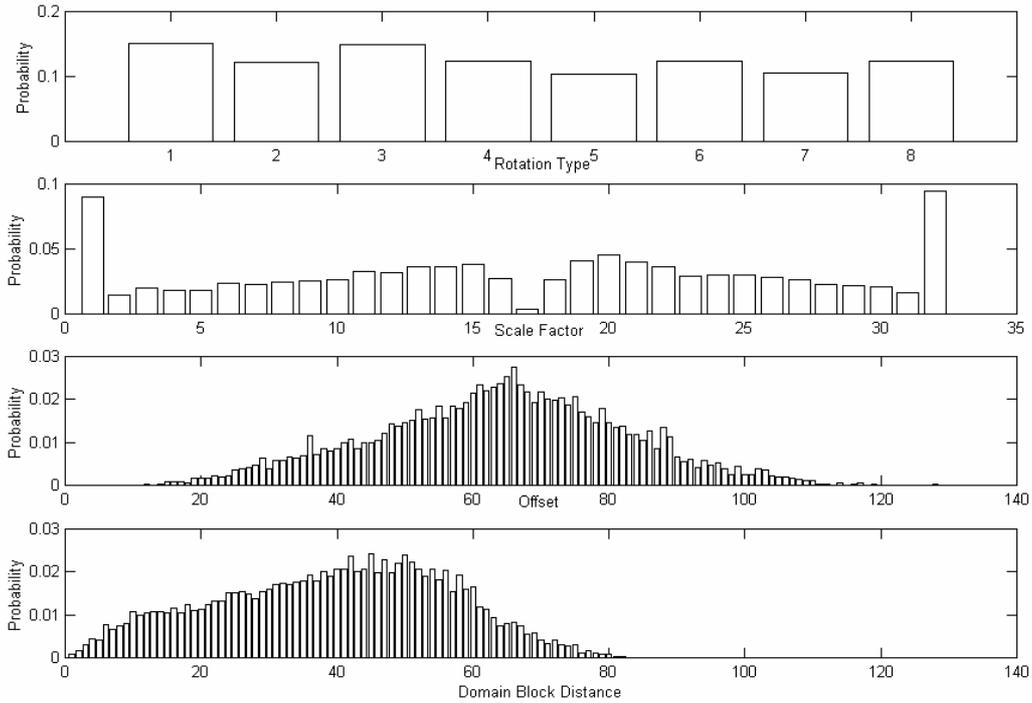

**Fig. 1.** The probability distribution of fractal parameters in image coding

As can be seen, the rotation parameter $\ell_i$ is near to uniform distribution, the scale parameter $\alpha_i$ is similar to the uniform distribution with peak values in the border, and the luminance offset $\Delta g_i$ and domain block position $D_{xy}$ are similar to normal distribution. According to the statistics, the domain block position $D_{xy}$ only happens in part of the space, and thus, encrypting $D_{xy}$ may waste some computing cost. For the scale parameter $\alpha_i$, it happens more frequently in the border than in other space. To encrypt $\alpha_i$, the chained encryption mode [33] should be taken to make the encrypted parameter in uniform distribution.

**3.2.3 Parameter Sensitivity**

The parameter-sensitivity of fractal encoding may be explained as follows: the decoding process is an iterated process, which causes that the reconstruction of each range block to depend not only on the matched domain block but also on other domain blocks. So the mistake of one decoded range block may lead to mistakes of other decoded range blocks, and after iterating the fractal transforms the decoded image may become unintelligible. Here, we define the fractal parameter sensitivity as the change rate of the image's quality degradation corresponding to different number of parameter mistakes. There is still no suitable measure for image's intelligibility. Since Peak Signal Noise Ratio (PSNR) is now often used to evaluate image quality, it is used here to measure image quality degradation. For 256-color images, the PSNR is defined as

$$PSNR = 10\log_{10}\frac{255^2}{MSE} \tag{5}$$

Where MSE (Mean Square Error) satisfies

$$MSE = \frac{1}{WH}\sum_{i=0}^{H-1}\sum_{j=0}^{W-1}[X(i,j) - X'(i,j)]^2. \tag{6}$$

Here, W and H are the image's width and height, respectively, and $X(i,j)$ and $X'(i,j)$ are the (i,j)-position pixels in the original image and the error-decoded image, respectively.

Experiments on various images (Lena, Boats, San, Couple and Eltoro) have been done, and the results are shown in Fig. 2. Here, the number of range block is 64×64, $\alpha_i$ is of 5 bits, $\Delta g_i$ is of 7 bits, $\ell_i$ is of 3 bits and $D_{xy}$ is of 10 bits. The curve corresponding to each parameter denotes the mean value of the test images. As can be seen, the curves of luminance shift $\Delta g_i$ and contrast scaling $\alpha_i$ are much lower than the others, which mean that $\Delta g_i$ and $\alpha_i$ cause much more quality degradation than other parameters do. Thus, the parameters, $\Delta g_i$ and $\alpha_i$, have higher parameter sensitivity than others, and they are prefer to be encrypted.

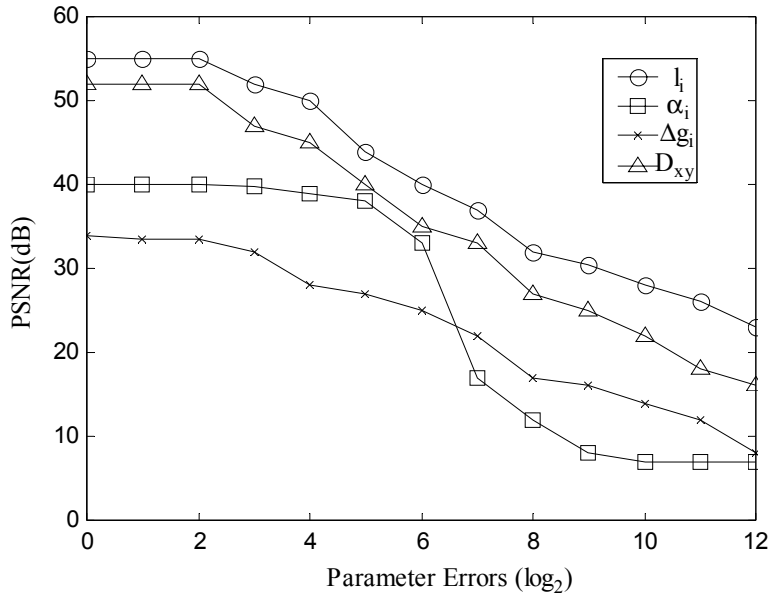

**Fig. 2.** Parameter sensitivity of fractal parameters

**3.3 Architecture of the Secure Coding Scheme**

**3.3.1 The Secure Encoding and Decoding Scheme**

According to the secure partial encryption principles described in Section 3.1 and the parameter investigation given in Section 3.2, the two parameters, i.e., luminance offset $\Delta g_i$ and contrast scaling $\alpha_i$, are more suitable for encryption than the isometric transform $\ell_i$ and domain block position $D_{xy}$. We propose the secure fractal encoding and decoding scheme based on parameter encryption, as shown in Fig. 3. In encoding, the image is partitioned into blocks, the fractal transform for each block is determined, the two parameters, i.e., $\Delta g_i$ and $\alpha_i$, are encrypted, and all the parameters are multiplexed and stored. In decoding, the parameters are de-multiplexed, the two parameters, i.e., $\Delta g_i$ and $\alpha_i$, are decrypted, and all the parameters are used to drive the iterated fractal transforms.

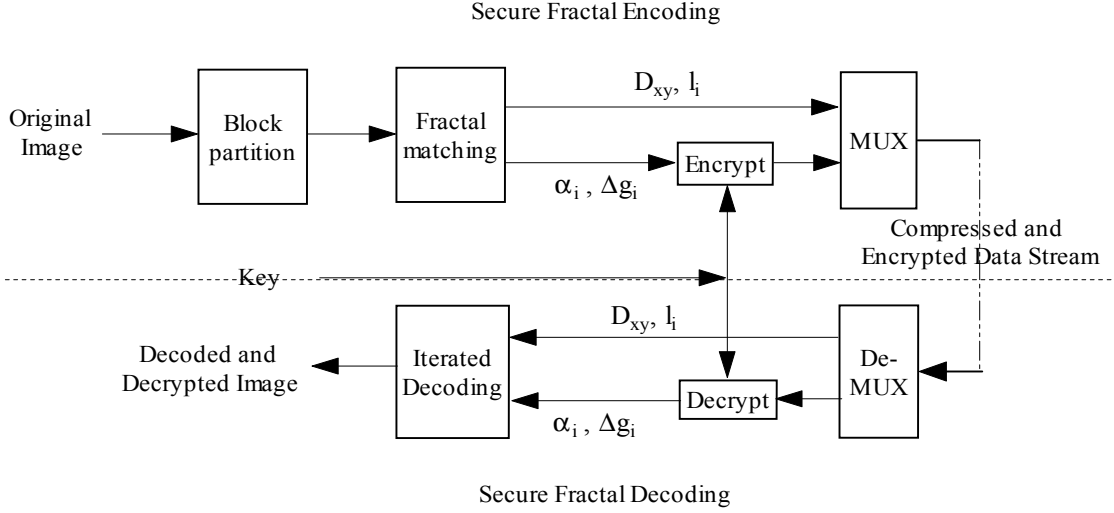

Secure Fractal Decoding

**Fig. 3.** The proposed secure fractal image encoding/decoding scheme

### 3.3.2 Parameter Encryption/Decryption

To improve the security of parameter encryption, the chained encryption mode is adopted. For N range blocks, N parameter pairs are generated, they are $X_0, X_1, \ldots, X_{N-1}$. Here, $X_i$ is the multiplex of the two parameters $\alpha_i$ and $\Delta g_i$, that is, $X_i = \alpha_i | \Delta g_i$ (i=0,1,…,N-1). Thus, the parameter encryption based on the chained encryption [33], as shown in Fig. 4, is defined as

$$\begin{cases} X_0' = E(X_0 \oplus V_0, K_0) = E((\alpha_0 | \Delta g_0) \oplus V_0, K_0) \\ X_{i+1}' = E(X_{i+1} \oplus X_i', K_{i+1}) = E((\alpha_{i+1} | \Delta g_{i+1}) \oplus X_i', K_{i+1}) \end{cases} \quad (7)$$

Here, $V_0$ is an initial vector, $K_i$ (i=0,1,…,N-1) is the i-th range block's encryption key, $X'_i$ (i=0,1,…,N-1) is the i-th encrypted range block, '$\oplus$' is the bitwise exclusive or operation, and E() is the encryption function. According to the size of $X_i$, the encryption function may be a block cipher or stream cipher.

In parameter decryption, the encrypted N parameter pairs $X'_0, X'_1, \ldots, X'_{N-1}$ are decrypted into $X_0, X_1, \ldots, X_{N-1}$ with the chained mode. Taking symmetric encryption for example, the decryption key $K_i$ (i=0,1,…,N-1) is same to the encryption key, as shown in Fig. 5. The decryption process is defined as

$$\begin{cases} D(X_0', K_0) \oplus V_0 = ((\alpha_0 | \Delta g_0) \oplus V_0) \oplus V_0 = \alpha_0 | \Delta g_0 = X_0 \\ D(X_{i+1}', K_{i+1}) \oplus X_i' = ((\alpha_{i+1} | \Delta g_{i+1}) \oplus X_i') \oplus X_i' = \alpha_{i+1} | \Delta g_{i+1} = X_{i+1} \end{cases} \quad (8)$$

Here, D() is the decryption function, and the other parameters are same to the ones in the encryption function. According to the size of $X_i$, the decryption function may be a block cipher or stream cipher, which is symmetric to the encryption function.

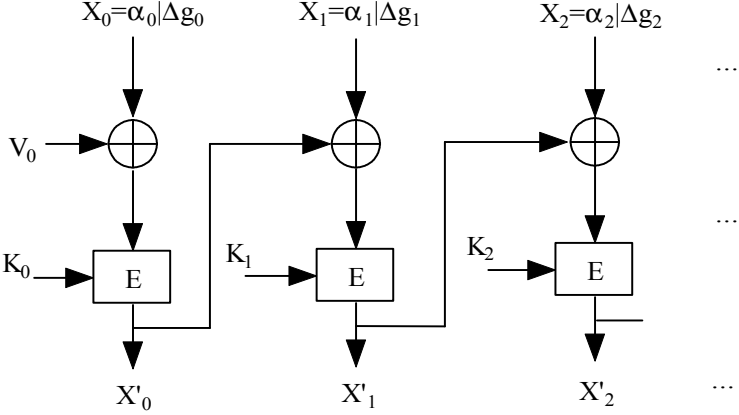

**Fig. 4.** The chained encryption mode for parameter encryption

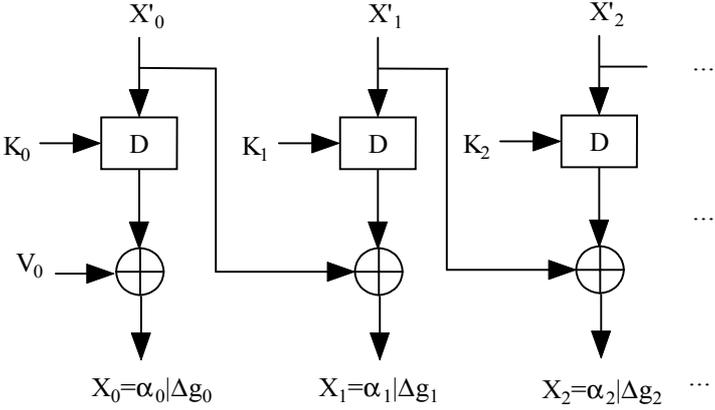

**Fig. 5.** The chained decryption mode for parameter decryption

## 4 Performance Evaluation

### 4.1 Security

For image encryption, the security depends on two aspects [2,34], i.e., cryptographic security and perceptual security. The former one denotes the encryption algorithm's security against such cryptographic attacks as brute-force attack, statistical attack, differential attack, etc. The latter one denotes the unintelligibility of the encrypted image content.

**4.1.1 Cryptographic Security**

In the proposed image encryption scheme, the selected parameters are encrypted by the cipher in the chained encryption mode. For cryptographic attackers, two points are often considered, i.e., counterfeit the parameters and break the cipher.

To counterfeit the parameters, $\alpha_i$ and $\Delta g_i$, the attackers' brute-force space is

$$B(N) = (2^6 \cdot 2^9)^N = 2^{15N}. \tag{9}$$

Here, N is the number of range block, the parameter space of $\alpha_i$ is $2^6$, and the one of $\Delta g_i$ is $2^9$. Generally, N is no smaller than 16x16, and thus, the brute-force space is no smaller than $2^{3840}$ that is too large for the attackers to get the original image.

The scheme's security against cryptographic attack depends mainly on the cipher. To keep secure, some existing ciphers with high security can be adopted, such as the block ciphers, Triple Data Encryption Standard (3DES) and Advanced Encryption Standard (AES), or the stream ciphers, Rabbit and RC4 [33]. Here, RC4 is recommended, since its security against some existing attacks has been evaluated and it is proved secure enough for various applications. In the following experiments, RC4 is used to encrypt the selected parameters with the key length of 128, and the same key is used for different range blocks.

**4.1.2 Perceptual Security**

In partial encryption, the perceptual security depends on the sensitivity of the encrypted parameters. In the proposed scheme, the fractal parameters with high sensitivity are encrypted, which keeps the encrypted images unintelligible. To compare the parameter sensitivity, different fractal parameters are encrypted, and the encrypted images are shown in Fig. 6 and Fig. 7. Here, the number of range block is 128×128, the number of domain block is 127×127, $\alpha_i$ is of 5-bit, $\Delta g_i$ is of 7-bit, $D_{xy}$ is of 14-bit, and $\ell_i$ is of 3-bit. As can be seen, if only the domain block position $D_{xy}$ is encrypted, the encrypted images (Fig. 6(d) and Fig. 7(d)) are still intelligible. By encrypting both the scale $\alpha_i$ and the offset $\Delta g_i$, the encrypted images (Fig. 6(e) and Fig. 7(e)) are both too chaotic to be understood. Furthermore, if the fractal image coding is based on quadtree, the parameter of quadtree height can also be encrypted, which produces also unintelligible images (Fig. 6(f) and Fig. 7(f)). The experiments on some other images get the similar results. Thus, the proposed encryption scheme can obtain high perceptual security.

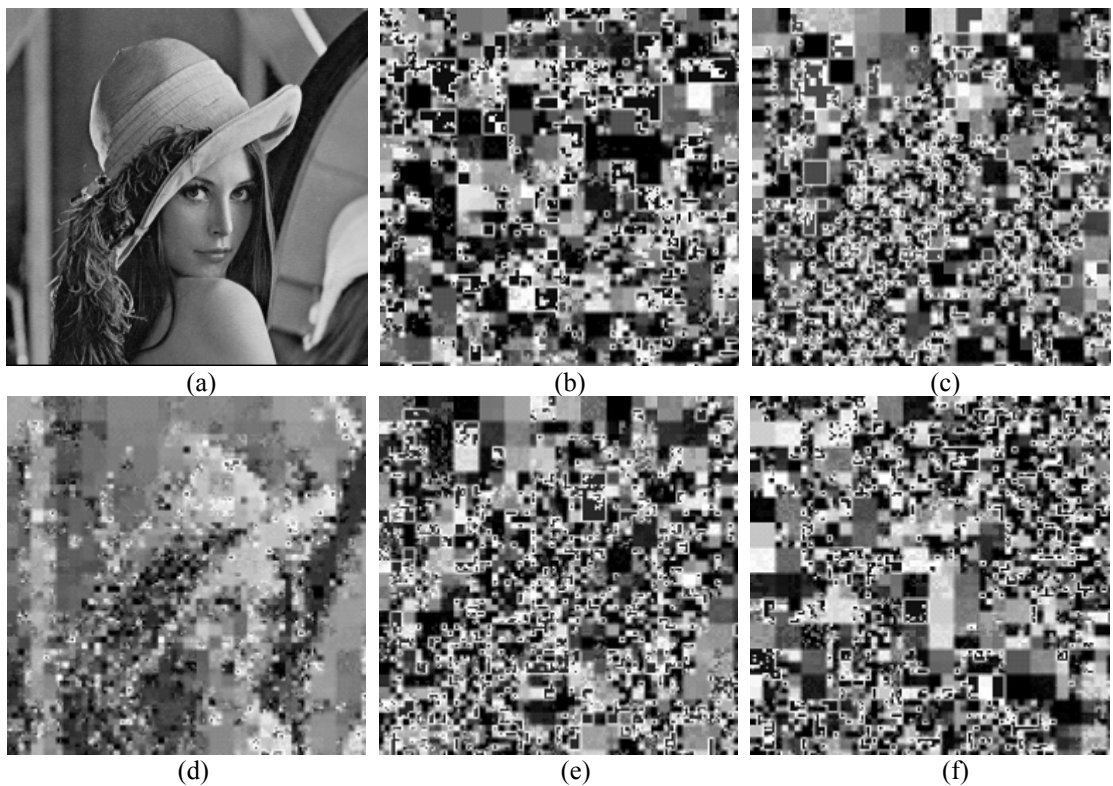

**Fig. 6.** The images produced by encrypting different parameters in Lena
((a)original, (b) encrypting the contrast scaling, (c) encrypting the luminance offset, (d) encrypting the domain block position, (e) encrypting both the contrast scaling and luminance offset, and (f) encrypting the contrast scaling, luminance offset and quadtree height.)

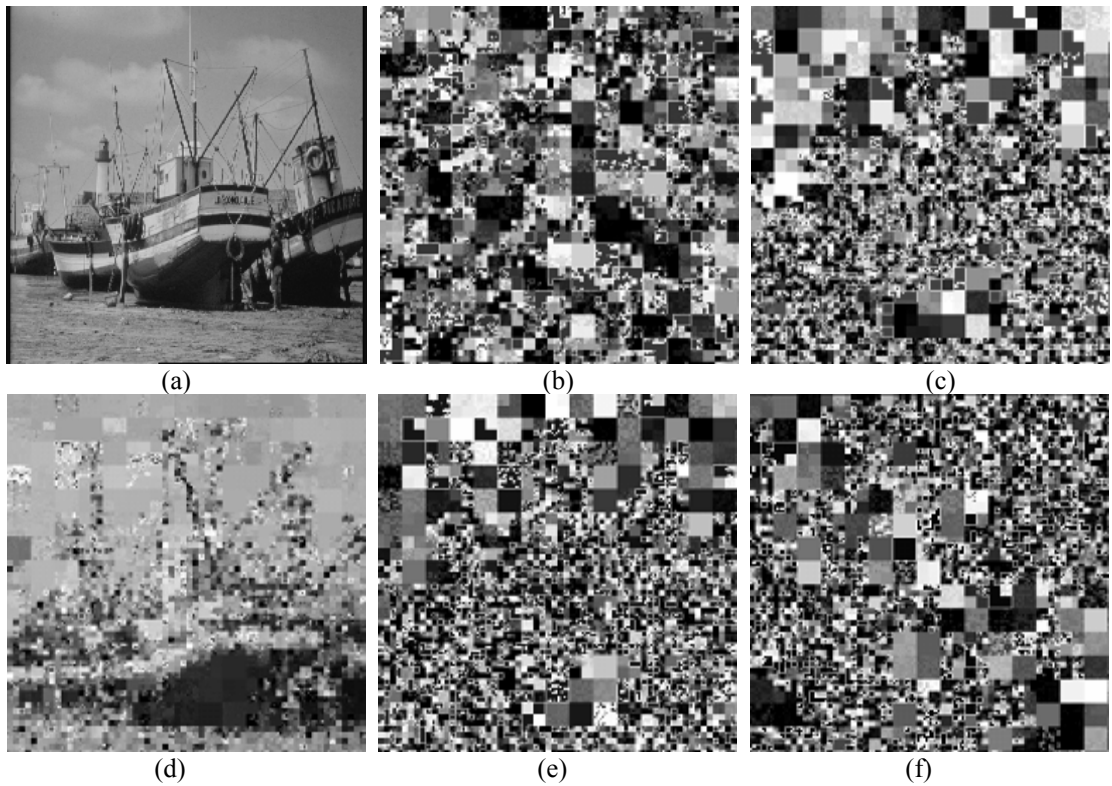

**Fig. 7.** The images produced by encrypting different parameters in Boats
((a)original, (b) encrypting the contrast scaling, (c) encrypting the luminance offset, (d) encrypting the domain block position, (e) encrypting both the contrast scaling and luminance offset, and (f) encrypting the contrast scaling, luminance offset and quadtree height.)

**4.1.3 Security against Some other Attacks**

Some attackers attempt to break the encryption scheme by making use of image data's properties. For example, they use different key to encrypt or decrypt the same image, and guess the key by investigation the difference between the encrypted or decrypted images. To counter this kind of attacks, high key sensitivity is required, which means that a slight difference in the decryption key will produce a quite different image. In the proposed scheme, the key sensitivity depends on the parameter sensitivity and the cipher's key sensitivity. Since the parameters with high sensitivity are selected, and the cipher with high security is adopted, the encryption scheme's key sensitivity can be confirmed. Taking Lena in Section 4.1.2 for example, it is firstly encrypted with the key '0123 4567 890A BCDE 0123 4567 890A BCDE' (Hex form), then decrypted with the key '0123 4567 890A BCDE 0123 4567 890A BCDF', '0123 4567 890A BCDE 0123 4567 890A BCDD' or '0123 4567 890A BCDE 0123 4567 890A BCDE', respectively. The results are shown in Fig. 8. As can be seen, the image can not be decrypted although there is only a slight change in the key. The experiments on some other images can get the similar result, which show that the encryption scheme is of high key sensitivity.

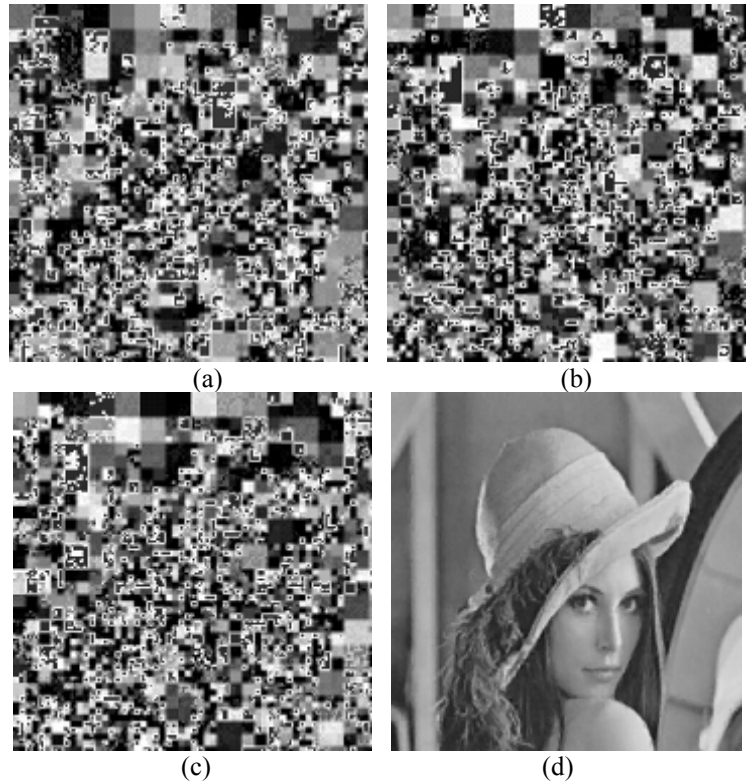

**Fig. 8.** Image encryption and decryption with different keys
((a) encrypted with Key='0123 4567 890A BCDE 0123 4567 890A BCDE' (Hex from) (b) decrypted with Key='0123 4567 890A BCDE 0123 4567 890A BCDF', (c) decrypted with Key='0123 4567 890A BCDE 0123 4567 890A BCDD', and (d) decrypted with Key='0123 4567 890A BCDE 0123 4567 890A BCDE')

**4.2 Encryption/Decryption Efficiency**

We have also tested the encryption/decryption speed of the proposed partial encryption scheme, which is measured by the encryption time ratio (Etr) and decryption time ratio (Dtr). Here, Etr and Dtr are defined as

$$\begin{cases} \text{Etr} = \dfrac{T_{\text{Encrypt}}}{T_{\text{Encode}}} \times 100\% \\ \text{Dtr} = \dfrac{T_{\text{Decrypt}}}{T_{\text{Decode}}} \times 100\% \end{cases}, \qquad (10)$$

where $T_{Encrypt}$, $T_{Encode}$, $T_{Decrypt}$ and $T_{Decode}$ denote the time cost of the encryption, encoding, decryption and decoding processes, respectively. Taking images of different sizes for example, the experimental results are shown in Table 4, where the parameters are the same as the ones in Section 4.1.2, and the computer is of 550MHz CPU and 128M RAM. Seen from Table 4, the encryption/decryption time is no more than 15% of the compression/decompression time. And the bigger the image is, the lower the time ratio is, which means that the encryption/decryption operation does not delay the compression/decompression process significantly. This property makes it easy to realize real-time compression and encryption and suitable for secure image encoding and transmission.

Table 4. Test of encryption/decryption speed

| Image (Size) | Encryption/Decryption Time Ratio | |
|---|---|---|
| | Encryption time ratio (%) | Decryption time ratio (%) |
| Lena(128×128) | 8.89 | 12.16 |
| San(128×128) | 8.42 | 12.15 |
| Couple(128×128) | 9.78 | 13.06 |
| Lena(256×256) | 5.43 | 7.07 |
| Boat(256×256) | 4.86 | 7.72 |
| San(512×512) | 2.02 | 3.41 |
| Eltoro(512×512) | 2.15 | 4.38 |

5 Conclusions and Future Work

In this paper, a secure fractal image coding scheme is proposed and evaluated. In this scheme, some parameters with large space and high sensitivity are encrypted during fractal image encoding and decrypted during fractal image decoding. The encryption or decryption operation combines with encoding or decoding operation and keeps the image format unchanged. The selection of parameters obeys the secure partial encryption principles, and the encryption scheme is proved secure by the experiments. Additionally, the encryption/decryption operation is time efficient compared with encoding/decoding operation. These properties make the encryption scheme suitable for secure image encoding or transmission. Note that, only the general fractal image coding is investigated in this paper, which will be extended to some typical fractal codecs in future work, such as the fractal image coding based on wavelet or DCT.